\documentclass[12pt,preprint]{aastex}
\usepackage{graphicx}
\usepackage[small,bf,up]{caption}

\newcommand{\CIV}{C{\sc iv}}
\newcommand{\MgII}{Mg{\sc ii}}
\newcommand{\ZnII}{Zn{\sc ii}}
\newcommand{\FeII}{Fe{\sc ii}}
\newcommand{\CaII}{Ca{\sc ii}}

\newcommand{\etal}{\textit{et al.\thinspace}}

\newcommand{\kms}{km s$^{-1}$}
\newcommand{\cm}{cm$^{-2}$}

\newcommand{\lsim}{\ \raise -2.truept\hbox{\rlap{\hbox{$\sim$}}\raise5.truept
        \hbox{$<$}\ }}
\newcommand{\gsim}{\ \raise -2.truept\hbox{\rlap{\hbox{$\sim$}}\raise5.truept
        \hbox{$>$}\ }} 
     
\def\toprule{\hline\hline}

\slugcomment{}

\shorttitle{Metal Systems in GRB/QSO sight-lines}
\shortauthors{Sudilovsky \etal }

\begin{document}

\title{Intervening Metal Systems in GRB and QSO sight-lines: \\ The \MgII\ and \CIV\ Question\altaffilmark{1}}
\altaffiltext{1}{Based on observations collected at the European Southern Observatory, Chile; proposal no. 75.A-0385, 75.A-0603, 77.D-0661}

\author{Vladimir Sudilovsky\altaffilmark{2,3}} %%%AUTHORS
\author{Sandra Savaglio\altaffilmark{3}}
\author{Paul Vreeswijk\altaffilmark{4}}
\author{C\'edric Ledoux\altaffilmark{4}}
\author{Alain Smette\altaffilmark{4}}
\author{Jochen Greiner\altaffilmark{3}}

\altaffiltext{2}{Physics Department, Guilford College, Greensboro, North Carolina} %%%AFFILIATIONS
\altaffiltext{3}{Max-Planck-Institut f\"ur Extraterrestrische Physik, Giessenbachstrasse, D-85748 Garching bei M\"unchen, Germany}
\altaffiltext{4}{European Southern Observatory,  Alonso de C\'ordova 3107, Casilla 19001, Santiago 19, Chile}

\begin{abstract} %%%ABSTRACT
Prochter \etal\ 2006 recently found that the number density of strong intervening $0.5<z<2$ \MgII\ absorbers detected in gamma-ray burst (GRB) afterglow spectra is nearly 4 times larger than in QSO spectra.  We have conducted a similar study using \CIV\ absorbers. Our \CIV\ sample, consisting of a total of 19 systems, is drawn from 3 high resolution and high to moderate S/N VLT/UVES spectra of 3 long-duration GRB afterglows, covering the redshift interval $1.6 < z < 3.1$. The column density distribution and number density of this sample do not show any statistical difference with the same quantities measured in QSO spectra. We discuss several possibilities for the discrepancy between \CIV\ and \MgII\ absorbers and conclude that a higher dust extinction in the \MgII\ QSO samples studied up to now would give the most straightforward solution. However, this effect is only important for the strong \MgII\ absorbers. Regardless of the reasons for this discrepancy, this result confirms once more that GRBs can be used to detect a side of the universe that was unknown before, not necessarily connected with GRBs themselves, providing an alternative and fundamental investigative tool of the cosmic evolution of the universe.

\end{abstract}


%%%%%%%%%%%%%%%%%%%%%%%%%%%%%%%%%%%%
%%%BEGIN MAIN TEXT
%%%%%%%%%%%%%%%%%%%%%%%%%%%%%%%%%%%%

\section{Introduction} \label{introduction} %%%INTRODUCTION%%%%%%%%%%%%%%%%%%
The advent of Gamma-ray burst (GRB) exploration in the last 10 years has changed our view of the universe. These highly energetic events have been found over a very large interval of redshift, from the local to $z = 6.3$ (Kawai \etal\ 2006). They are so bright that when one of these events occurs, the most remote structures can be temporarily ``illuminated'' and studied in unprecedented detail (Vreeswijk \etal\ 2005; Chen \etal\ 2005). At $z>1.5$, the optical (UV rest frame) afterglows of long duration GRBs are revealing properties of the forming universe never detected before, such as the existence of a population of metal enriched, star-forming and relatively small galaxies (Berger \etal\ 2006; D'Elia \etal\ 2006; Fynbo \etal\ 2006; Prochaska \etal\ 2006; Savaglio \etal\ 2006).

In a recent work, Prochter \etal\ (2006; hereafter P06) used GRBs to probe the galaxy halos distributed along GRB sight-lines, similarly to what has been done for decades using QSOs. In their study, they identified 14 strong (equivalent width larger than 1 \AA) intervening \MgII\ absorbers in the sight-lines of 14 long duration GRBs. The total path-length is $\Delta z = 15.5$ and the mean redshift is $\langle$z$\rangle$ = 1.1, for a mean number density $dn/dz=0.90\pm0.24$. Surprisingly, only an incidence of 3.8 similar \MgII\ absorbers is expected for quasars (QSOs) of the same path-length. Dust extinction, gravitational lensing, different beam sizes, and systems belonging intrinsically to the GRB source were ruled out as possible explanations for this difference. Following such a result, Porciani \etal (2007) proposed that the different behavior of \MgII\ in GRB and QSO sight-lines might be a combination of dust extinction bias, gravitational lensing, and absorbers being misidentified as intervening.

Particularly intriguing is the recent discovery of absorption variability in $z=1.48$ \MgII\ and \FeII\ intervening systems along  GRB~060206 sight-line, over a time interval of a few hours  (Hao \etal\ 2007). These authors support a scenario according to which the variability is expected if the size of the absorbing clouds is of the order of the GRB beam size (i.e.\ $10^{16}$ \cm) that expands 
faster than light (Loeb \& Perna 1998), due to the large Lorentz factor. In addition, micro-lensing could also add to the possible cause of variability in this case, as described  by Lewis \& Ibata (2003).

In an effort to understand the reason of the \MgII\ discrepancy, we have considered intervening \CIV\ absorbers in the lines of sight to GRB and QSO sources. It is important to note that the aforementioned effects act in a different way in \CIV\ intervening systems in GRB and QSO sight-lines. For instance,  \CIV\ is found at much larger impact parameters relative to \MgII, making gravitational lensing less effective. This also implies that the strength of \MgII\ absorption does not correlate to an increase in the strength of \CIV\ absorption (York et al.\ 2006). Moreover, \CIV\ absorbers do not trace dust to nearly the same extent as \MgII\ absorbers. Effects from a partial covering factor of QSO background sources (due to a large QSO beam size; Frank \etal\ 2007) are ruled out for both \CIV\ and \MgII\ absorbers, as this was never detected in high-resolution and reasonable S/N spectra.

Both \MgII\ and \CIV\ are easily detected due to their relatively high cosmic abundance, large oscillator strength, and their appearance as doublets. \CIV\ absorbers, a highly ionized species\footnote{The ionization potential of \CIV\ is 64.5 eV, compared to  15.0 eV for \MgII .}, are common features in the large galactic halo (Chen \etal\ 2001), while strong \MgII\ absorbers are found at low impact parameters  ($\rho \approx$ 15 kpc; Lanzetta \etal\ 1990). Strong \MgII\ systems with a rest equivalent width (W$_{r}$) in excess of 0.6 \AA\ are often associated with damped Lyman-$\alpha$ (DLA) systems (Rao \etal\ 2006; York \etal\ 2006). A direct comparison between the properties of \CIV\ and \MgII\ absorbers is complicated by the fact that \CIV\  can be effectively detected in the optical for $z > 1.8$, while \MgII\ is detected at lower redshifts ($0.2 < z < 2.3$). 

Our \CIV\ and \MgII\ survey in GRBs include optical spectra of 5 GRB afterglows taken with the high resolution (R $>$ 42000) UV-optical echelle spectrograph UVES at VLT (Vreeswijk \etal\ 2007). \CIV\ absorbers in 3 sight-lines are compared with those found in QSOs that cover very similar redshifts (Boksenberg \etal\ 2003). The cumulative number of absorbers as a function of redshift as well as the column density distribution are used to compare the incidence of \CIV\ systems in sight-lines of the GRBs and QSOs. Our \MgII\ sample is used to confirm the result obtained with equivalent widths.

In \S \ref{thedata}, the data are presented. The methods for the analysis are presented in \S \ref{methods}, and the result and discussion are found in \S \ref{results}.. Throughout the paper we adopt a $h \equiv H_o/100= 0.7$, $\Omega_M = 0.3$, $\Omega_\Lambda = 0.7$ cosmology (Spergel \etal\ 2006).

%%%%%%%%%%%%%%%%%%%%%%%%%%%%%%%%%%%%
\section{The Data} \label{thedata} %%%THE DATA
\begin{figure}[htbp]%%%FIGURE: VELOCITY PROFILES
   	\begin{centering}
    	\includegraphics[scale = .22]{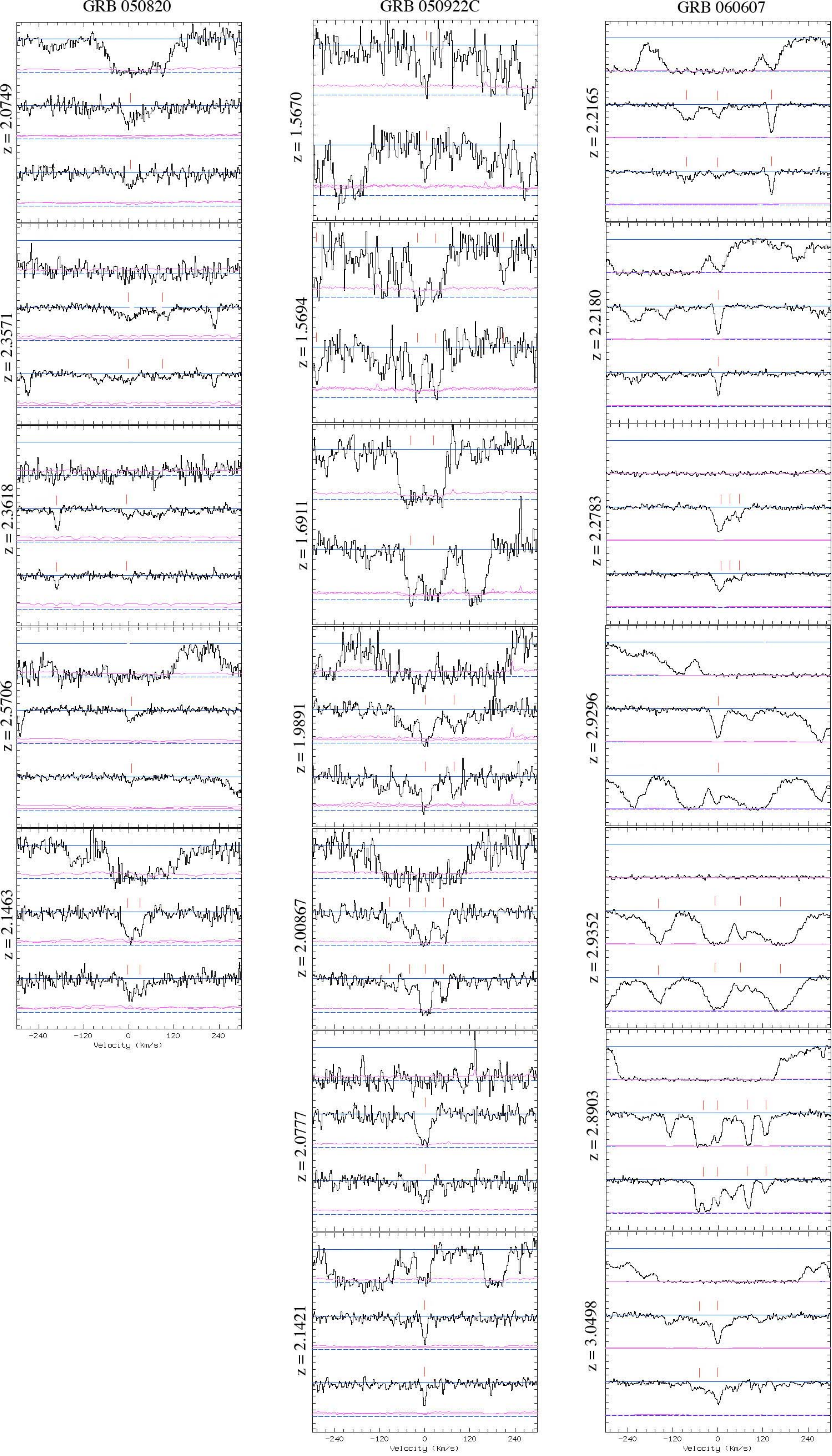}
	\caption{\small Normalized flux vs Velocity profiles for \CIV\ systems in the three GRBs studied. From left to right are GRB 050820, GRB 050922C, and GRB 060607. The pink line represents the associated error spectra. The panels, from top to bottom, plot Ly-$\alpha$ (when applicable), \CIV 1548, and \CIV 1550.}
    	\label{fig:VelocityProfiles}
	\end{centering}
    	\end{figure}

GRB optical afterglow spectra of 5 GRBs were taken using the high resolution spectrograph UVES. These are: GRB 050730 at $z$ =  3.967 (Chen \etal\ 2005b; D'Elia \etal\ 2007), GRB 050820 at $z$ = 2.612 (Ledoux \etal\ 2005; Prochaska \etal\ 2005; ), GRB 050922C at $z$ = 2.198 (Jakobsson \etal\ 2005; Jakobsson \etal\ 2006; Piranomonte \etal\ 2007), GRB 060418 at $z$ = 1.489 (Dupree \etal\ 2006; Vreeswijk \etal\ 2007), and GRB 060607 at $z$ = 3.082 (Ledoux \etal\ 2006). The spectra were normalized by fitting a polynomial function to the  continuum. The spectra of GRBs 050820, 050922C, and 060607 were used for the \CIV\ sample with a total absorption path-length of $\Delta z = 2.25$ and a mean redshift of $\langle z \rangle = 2.63$. The two other GRB spectra  -- GRB 060418, 050730 -- were excluded from the sample, due to lack of \CIV\ redshift coverage and poor S/N. For the \MgII\ analysis, we used all these spectra, as the EWs are robust measurements when the S/N is not very high. The spectra used in the \CIV\ survey all have high S/N and resolution (Table \ref{table:samplecharacteristics} and Fig.~\ref{fig:VelocityProfiles}).

The QSO \CIV\ sample (Boksenberg \etal\ 2003) has a very similar $\Delta z$ and $\langle z \rangle$ of 2.26 and 2.65, respectively. A total of 20 \CIV\ systems were identified in the GRB sample, and a total of 40 systems in the QSO sample. The factor of two difference in the number of systems is due to the higher S/N and thus lower detection limits of the QSO sample. Information specific to particular GRBs or QSOs are found in Table \ref{table:samplecharacteristics}.

The software ESO-MIDAS and the context FITLYMAN were used to derive the column density, $N$, by simultaneously fitting Voigt profiles to both lines of the \CIV\ doublet. If a system contained multiple components, the column density used was the sum of the column densities of the components.

\subsection{\CIV\ Selection techniques} \label{selectiontechniques}%%subsection: CIV SELECTION TECHNIQUES
The \CIV\ systems usually are well defined and have multiple absorption features with a velocity distribution structure of up to a few hundred \kms\ (Ellison \etal\ 2000; Boksenberg \etal\ 2003; Songaila 2005; Songaila 2006). One or more \CIV\ `cloudlets' can compose a \CIV\ system. Individual systems (i.e.\ belonging to one physical entity such as a galaxy) are usually separated from each other by more than a few hundred \kms, and are often associated with Ly-$\alpha$ absorption. In two cases, several \CIV\ absorption features that share the same Ly-$\alpha$ absorption, were classified as two distinct systems. All other \CIV\ absorption features associated with distinct Ly-$\alpha$ absorption are considered individual systems. The velocity spread of these \CIV\ absorption features would have been in excess of 600 \kms\ if classified as one system. The column density of the \CIV\ system is the sum of the column densities of all of its constituent `cloudlets'. Velocity profiles of all identified systems are found in Fig.~\ref{fig:VelocityProfiles}. Table \ref{table:system_characteristics} presents the redshift, column density $N_{\rm C IV}$, and Doppler parameter $b$ of all \CIV\ systems detected in the GRB sample

In this study, only \CIV\ \emph{systems} outside of the Lyman-$\alpha$ forest (i.e., $z \ge 1215.67\times(1 + z_{em})/1548.1949-1$) are included in the sample. No systems within $\Delta v = 2300$ \kms\ (corresponding to a physical separation of 5.5 Mpc at $z=2.3$) of the redshift of the GRB of QSO were included to prevent any unknown differential effects between the GRB and QSO host environment. A total of 3 \CIV\ systems were excluded from the QSO sample, and 1 system from the GRB sample.

\subsection{Completeness limit} \label{limits}%%subsection:COMPLETENESS LIMITS

Since the S/N in the spectra vary, we derived the lowest detection limit in our GRB sample. A completeness limit of 1 \AA\ rest equivalent width (W$_{r}$ $\ge$ 1 \AA) was used by P06 for the \MgII\ sample. In this study, column density is used as the parameter to determine completeness. A theoretical rest equivalent width limit of 3 $\sigma$ is derived from the various error spectra for the weaker line of the doublet (\CIV$\lambda1550$). Since the ratio of the stronger to the weaker line is 2:1, for small absorptions (i.e.\ in the linear part of the curve of growth), this corresponds to a detection limit of 6 $\sigma$ for the stronger line (\CIV$\lambda1548$). When combined with the fact that \CIV\ systems are identified by the presence of \emph{both} of these lines, our \CIV\ detection limit is conservative and is $>6 \sigma$. 

\begin{figure}[htbp]%%%FIGURE: LIMITS
   	\begin{centering}
    	\includegraphics[scale = .5]{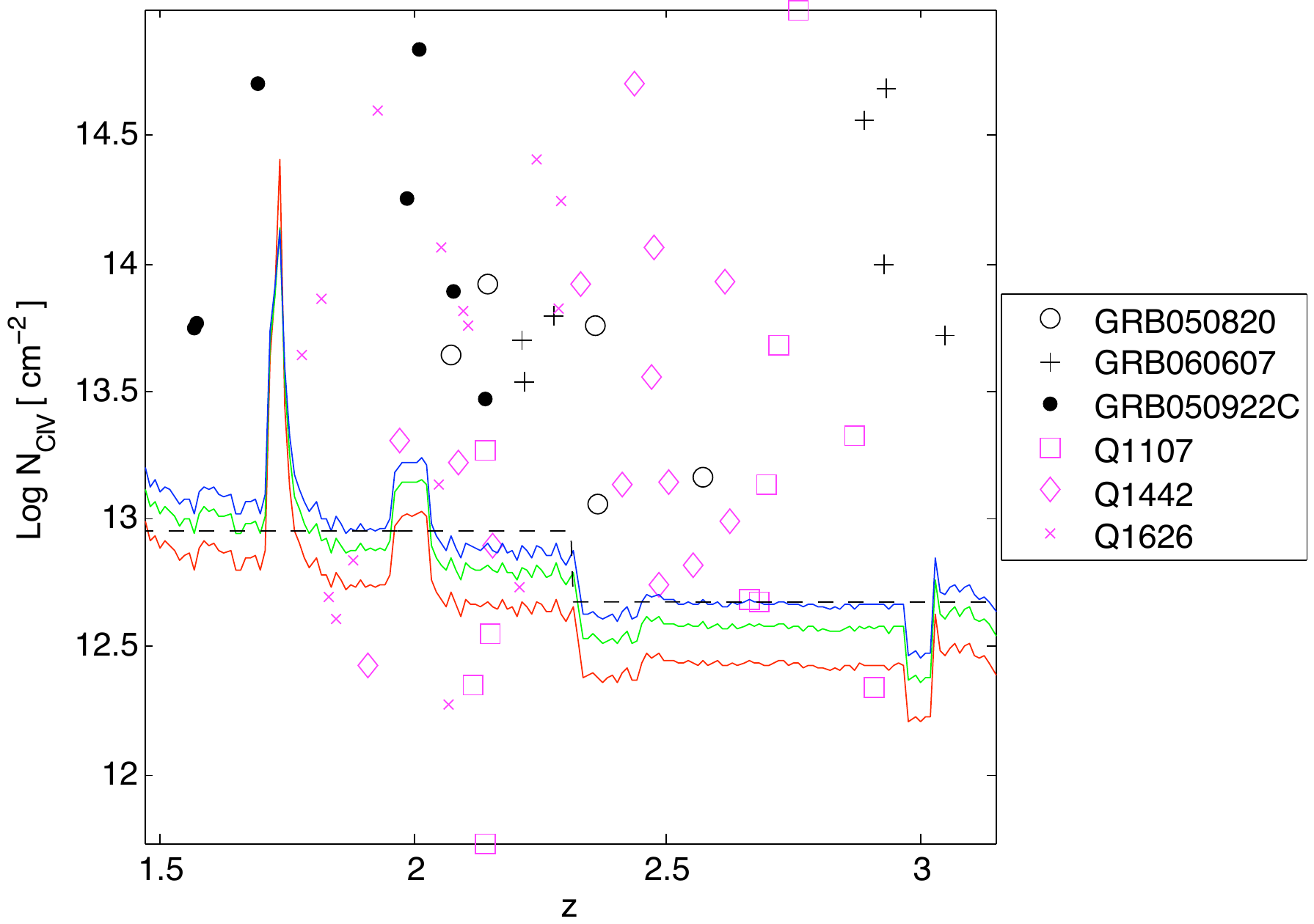}
	\caption{\small Column density of \CIV\ systems, measured in GRB (black points) and QSO (pink points) sight-lines, respectively, as a function of redshift. The lines represent the completeness limit to detect lines at $>6\sigma$ confidence level. From the bottom to top, the limits for $b=[5, 10, 15]$ \kms\ are plotted. The flat dashed line represents the limit used in this study.}
    	\label{fig:limits}
	\end{centering}
    	\end{figure}

This value is converted to a column density limit for a given doppler parameter $b$, from the linear portion of the empirical curve of growth of \CIV . The final column density limit for the survey is the maximum column density limit among the 6 spectra of the 3 GRBs (Fig.~\ref{fig:limits}). Limits were calculated using $b=[5,10,15]$ \kms\ (colored lines from bottom to top in Fig.~\ref{fig:limits}). We notice that the mean Doppler parameter in the sample is $\langle b \rangle = 14.5$ \kms, and that large $b$ values are typically associated with large \CIV\ column density, which are easy to detect.

As our S/N does not vary much for $z < 2.3$, and for $z > 2.3$ (apart from a couple of two small redshift bins) we have chosen for the final analysis, a conservative limit close to the  $b=15$ \kms\ line, described by a step function, with $N_{lim}=10^{13}$ cm$^{-2}$ and  $N_{lim}=10^{12.7}$ cm$^{-2}$, in the lower and higher redshift bin, respectively (dashed line in Fig.~\ref{fig:limits}).		
As the average resolutions of the GRB and QSO spectra are within 5\% of each other and are all well above the value to resolve the \CIV\ doublet, it is not necessary to compensate for different resolutions when comparing the different spectra (Table \ref{table:samplecharacteristics}).

The black and the pink points in Fig.~\ref{fig:limits} represent the measured GRB \CIV\ and QSO \CIV\ column densities, respectively. 

\subsection{The QSO sample} \label{QSOsample}%%subsection:THE QSO SAMPLE
Our QSO sample comes from a larger sample of quasars used by Boksenberg \etal (2003). We use their identification of \CIV\ systems and column densities, and we derived cumulative numbers and column density distribution from their tables. The resolution and redshift range of the QSO sample is very similar to the GRB sample. However, the S/N is significantly higher in the QSO sample (Table \ref{table:samplecharacteristics}).
		
%%%%%%%%%%%%%%%%%%%%%%%%%%%%%%%%%%%%%%%%%%%%%	
\section{Methods} \label{methods} %%%METHODS

\subsection{Mg~II absorbers in the GRB sight-lines}\label{MgII_absorbers} %%MgII ABSORBERS IN THE UVES sight-lineS

To verify if the over-density of \MgII\ systems along GRB sight-lines is
not due to the inhomogeneity of P06's sample, we have performed the same analysis on our homogeneous sample of UVES spectra using {\it exactly} the same search criteria. We notice that two UVES afterglow sight-lines are not in the P06's sample: GRB~050922C and GRB~060607. Table \ref{table:MgII_survey} shows the GRB sight-lines used, the starting and ending redshift of the potential \MgII\ system, the redshifts of the identified \MgII\ systems, and the rest-frame equivalent width that we measured. 

The total redshift path covered by the UVES sample is $dz$=6.75 with a mean redshift of $\langle z \rangle$ = 1.3.  We find 6 \MgII\ systems with a rest-frame equivalent width larger than
unity. This corresponds to a number density $dn/dz=0.89\pm0.36$, consistent with the  $dn/dz=0.90\pm0.24$ found by P06, at a mean redshift $\langle z \rangle = 1.1$. 

Although our redshift path is smaller than that of
P06, our sample of UVES spectra is homogeneous and of much
higher quality, and provides a preliminary confirmation of the \MgII\ excess observed in GRB sight-lines.  As the sample will grow, a more sophisticated
analysis will be possible, such as the one we present in the following
sections on the \CIV\ systems.

\subsection{ Cumulative number of \CIV\ systems vs z}%%subsection:CUMULATIVE NUMBER VS Z

\begin{figure}[htbp]%%%FIGURE: NUMBER DENSITY
   	\begin{centering}
    	\includegraphics[scale = .5]{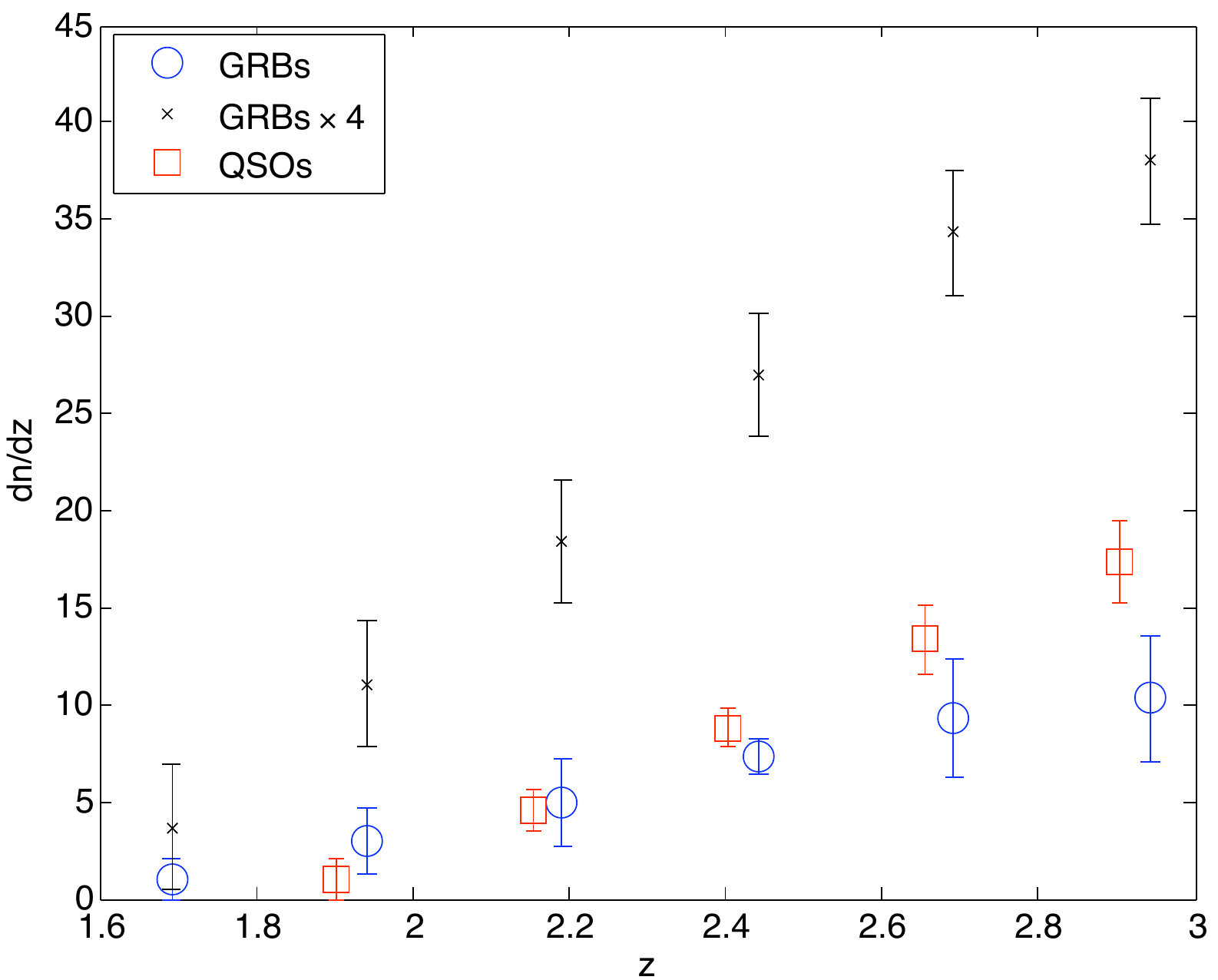}
	\caption{\small Cumulative number of \CIV\ systems in GRB (open circles) and QSO (open squares) sigthlines, per redshift bin of $\Delta z=0.25$, as a function of redshift. The data designated as GRBs $\times$ 4 (crosses) represents an incidence of 4 times as many \CIV\ systems as were actually measured.}
    	\label{fig:NumberDensity}
	\end{centering}
    	\end{figure}

The cumulative number of \CIV\ systems as a function of redshift is shown in Fig.~\ref{fig:NumberDensity} for the QSO and GRB samples. A bin size of $z = 0.25$ was used and error bars were calculated assuming a Poisson distribution, where $\sigma \propto \sqrt{n}$, with $n$ the number of \CIV\ systems. In addition to providing an easy visual comparison with the results obtained by P06 , this method has the potential to amplify systematic differences in the number of \CIV\ systems that are too small to notice on a redshift bin by bin basis. 

Fig.~\ref{fig:NumberDensity} shows no significant difference between the GRB and QSO samples. The detection limit discussed in \S \ref{limits} was applied to both the GRB and QSO subsamples. To demonstrate that an excess of $\sim4$ times detected in \MgII\ systems would be easily detected in our \CIV\ study, we multiply by four the cumulative number of \CIV\ systems measured in the GRB sample (plotted as crosses in Fig.~\ref{fig:NumberDensity}).

In addition, the data were split into two subsamples based on column densities, for $N_{CIV} > 10^{13.8}$ cm$^{-2}$ and $N_{CIV} \le 10^{13.8}$ cm$^{-2}$. This corresponds to the median column density of the full \CIV\ sample. Though in \S \ref{introduction} we mentioned that there has not been a relationship established between \CIV\ and \MgII\ systems, we split the sample to test if there is any difference in \CIV\ systems based on their absorption strength. Again, there is no detectable difference in the number density of \CIV\ systems for GRB or QSO sight-lines, although small number statistics play a more significant role for the subsamples than in the full sample.

%In Figs.~\ref{fig:NumberDensitySmallN} and \ref{fig:NumberDensityLargeN}, we show the results for $N_{\rm C IV}> 10^{13.8}$ cm$^{-2}$ and $N_{\rm C IV}\le 10^{13.8}$ cm$^{-2}$, respectively. A bin size of $\Delta z$ = 0.35 was used this time, in order to populate each bin.

\subsection{Column Density Distribution}%%subsection:COLUMN DENSITY DISTRIBUTION

The number of \CIV\ systems per column density interval and
absorption path-length $d^2n/dNdX$, in GRBs is compared with the same function derived from QSOs. The absorption path-length is defined as:
	
	\begin{equation}
		dX = \frac{1 + z}{\sqrt{\Omega_{M}(1 + z) + \Omega_{\Lambda}/(1 + z)^{2}}}dz ~.
	\label{eqn:dX}
	\end{equation}	

\noindent	 	 
We selected a column density interval of $dN = 10^{0.3}$ cm$^{-2}$, while $z \equiv \langle z \rangle = 2.37$, corresponds to the mean redshift of the sample. The redshift bin $dz$ is the total redshift interval covered by all sources, from $z_{min}$ to $z_{max}$, as described in \S \ref{selectiontechniques}.

Using the column density distribution to compare the incidence of \CIV\ systems allows one to account for multiple redshift coverage as well as compare systems in spectra with different S/N, without discarding any data due to different completeness limits. In addition, the column density distribution provides an easy comparison of the incidence of strong and weak absorbers alike in different lines of sight.

\begin{figure}[htbp] %%%FIGURE: COLUMN DENSITY DISTRIBUTION
   	\begin{centering}
    	\includegraphics[scale = .5]{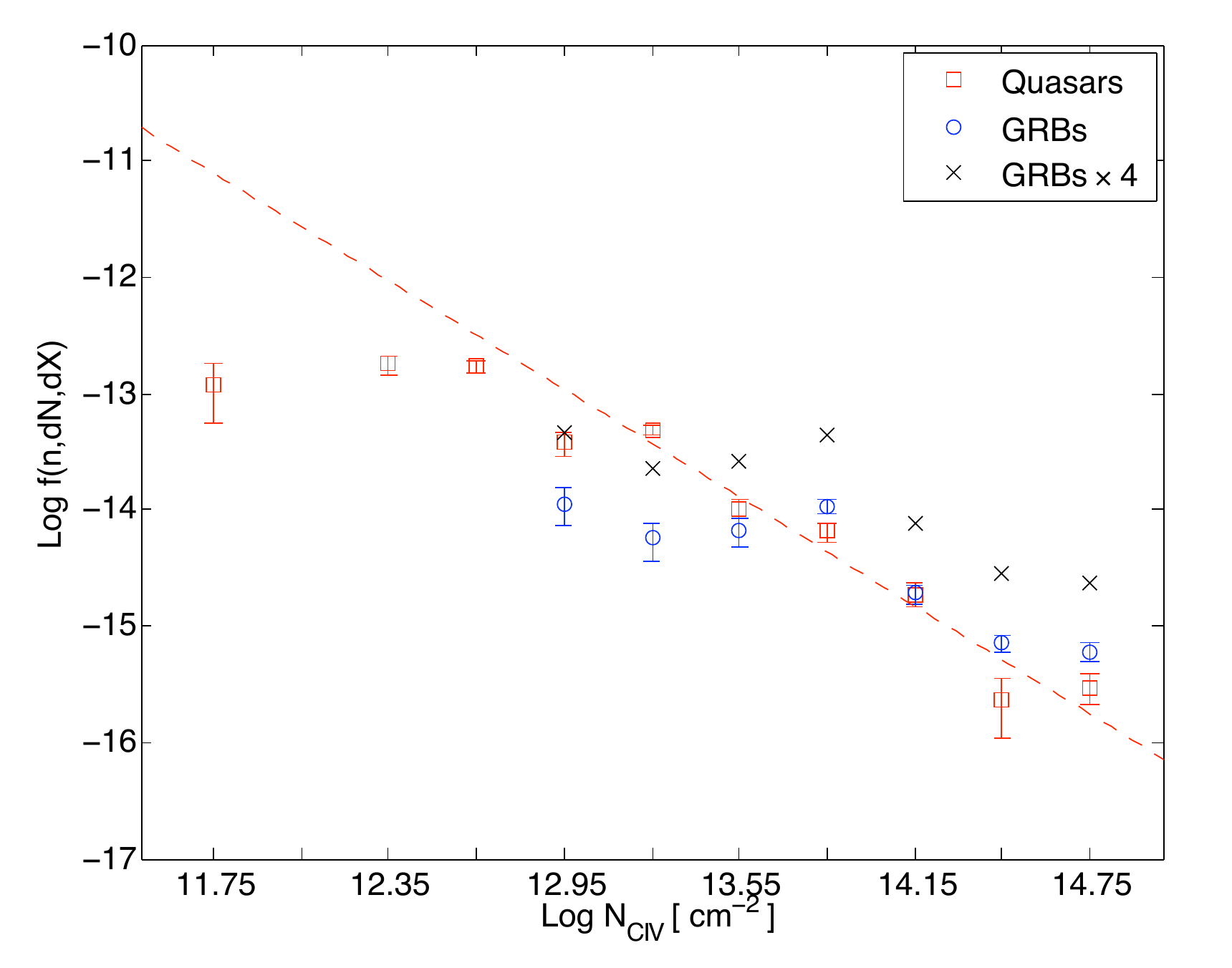}
	\caption{\small Column density distribution of the GRB (open circles) and QSO (open squares) samples. Plotted in black crosses is the column density distribution of 4 times as many \CIV\ systems than there actually were. The data are best fit by a line with a slope of [-1.5$\pm$0.2, -1.2$\pm$0.5] and offset of [6$\pm$3, 3$\pm$6] for the QSO and GRB sample, respectively. The best fit for the QSO sample is plotted as a dashed line.}
    	\label{fig:columndensitydistribution}
	\end{centering}
    	\end{figure}

In Fig.~\ref{fig:columndensitydistribution} we show the results. The sample completeness is demonstrated by the flattening of the power-law distribution, which happens at $\log N_{\rm C IV}=13.6$ and $\log N_{\rm C IV}=12.6$, for the GRB and QSO samples, respectively. The two column density distributions above these limits are best fit with a power-law of the form $dn/dN \propto N^{-\beta}$, where $\beta = 1.5 \pm 0.2$ for the QSOs, and $\beta = 1.2 \pm 0.5$ for the GRBs.

As in the case of the cumulative number, the \CIV\ column density distributions in GRB and QSO sight-lines are consistent. We also show that an excess of $\times 4$ would show a clear deviation of the two distributions.

%%%%%%%%%%%%%%%%%%%%%%%%%%%%%%%%%%%%

\section{Results} \label{results}%%%RESULTS

We have studied the statistical difference between intervening absorbers in GRB and QSO sight-lines. Our \MgII\ sample confirms the overdensity detected by P06 in GRB sight-lines. On the other hand, there is no statistical difference in the number of \CIV\ systems, considering both the cumulative number and the column density distribution. \MgII\ and \CIV\ do not behave the same way. We consider the effects of dust extinction, gravitational lensing, and absorbers associated with the circumburst environment in the following subsections. We note that the redshift intervals covered by the \CIV\ and \MgII\ absorbers are different, $1.6<z<3.1$ for the former and $0.5<z<2$ for the latter.

\subsection{Dust Extinction}

It has been suggested that QSO surveys are biased against strong absorbers distributed along the sight-lines because dust can obscure the background source (Fall \& Pei 1993; Smette \etal\ 2005 astro-ph/0504657). The dust extinction in \CIV\ absorbers is negligible. To estimate this, we consider that the typical column density in the observed GRB and QSO samples of intervening \CIV\ is $N_{\rm C IV}<10^{14.8}$ \cm . It is reasonable to believe that the carbon is marginally depleted in dust grains (Savage \& Sembach 1996). We assume that most of the carbon in these gas clouds is in \CIV\ form, and that the dust-to-metals ratio is as in the Large Magellanic Cloud (Bohlin \etal\ 1978). Following Savaglio \& Fall (2004), we estimate dust extinction by making the basic assumption that it is proportional to the total column density of metals.  A correction must be applied in the case of a refractory and/or ionized heavy element species is considered. However, no correction is necessary if the heavy element considered is not depleted in dust and the ionization correction is negligible. Using LMC dust-to-metals ratio and CIV column density yields an upper limit to the visual and UV  dust extinction of $A_{1500}=0.003$ and $A_V=0.001$, respectively, for $N_{CIV}<10^{14.8}$ \cm . One should recall that the LMC dust-to-metals ratio is generally small compared to other regions of the dense ISM in the Milky Way, and that CIV is a high-ionization species and thus a large ionization correction is necessary to get the total carbon column density. If carbon is only 10\% in \CIV\ form and the dust-to-metals ratio is 10 times higher than in the LMC (perhaps as in the Galactic cool ISM), the extinction would be 100 times larger, but still negligible, even in the UV.

The picture could be different for the \MgII\ absorbers selected by P06. The dust depletion of magnesium in the Galaxy is anywhere between 70\% to 95\%, from the warm diffuse to the cool diffuse ISM (Savage \& Sembach 1996). The selection by P06 of $W_r(2796)>1$ \AA\ corresponds to approximately a \MgII\ column density limit of $N_{MgII}\gsim 10^{14}$ \cm. For $N_{MgII}= 10^{14}$ \cm, the dust extinction is reasonably negligible, assuming a LMC extinction law. However, the dust extinction grows linearly with the metal column density, if the dust-to-metals ratio is constant (Savaglio \& Fall 2004). For $W_r(2796)>1.5$ \AA\  (half of the \MgII\ absorbers in P06's and the present
UVES samples) we derive $N_{MgII}\gsim 10^{15}$ \cm, and  thus $A_{1500}>0.07, 0.7$, in the case of warm diffuse or cool diffuse ISM, respectively.  

Though dust cannot explain the difference for the $W_r=1-1.5$ \AA\ absorbers, the detection can be more complicated for stronger absorbers with $W_r>1.5$ \AA, which is more than half of P06's sample. This is particularly true if these absorbers are characterized by a dust depletion as in the Galactic cool ISM.

There has been extensive debate on the topic of dust extinction in QSOs. Ellison \etal\ 2004 found no significant selection biases associated with the SDSS QSO data, in the CORALS survey. However the sample at $0.6 < z < 1.7$ includes all \MgII\ absorbers with $W_r>0.3$ \AA, which are mostly not affecting the background source. Indeed, Nestor \etal (2006) found that the column density distribution of weak \MgII\ absorbers is different than that for strong absorbers, which could suggest an evolving dust extinction bias with column density. York \etal (2006), found a negligible extinction of ($E_{B-V}$) $<$ 0.001 for sight-lines with \MgII\ systems with W$_{r} <$ 1.5 \AA, in agreement with our expectations. The extinction is slightly higher above this limit. Vladilo \&\ Peroux (2005) find that dust extinction becomes significant ($A_{V} \approx 0.15$ mag) as the column density of \ZnII\ approaches $10^{12.8}$ cm$^{-2}$, which in turn corresponds to a \MgII\ W$_{r}$ $\approx$ 3 \AA\ (Turnshek 2005; Vladilo 2005). Though  3 \AA\ is an unusually strong absorption, it is unclear at what \MgII\ EW between $W_r \approx 1.5$ \AA\ and $W_r \approx 3$ \AA\ does dust extinction start to affect surveys. Interestingly, Wild \etal\ (2005; 2006) found that strong \CaII\ absorbers at $0.8<z<1.3$ are missed by the SDSS, due to dust obscuration. The EW of \MgII\ associated with these CaII absorbers is typically above 2 \AA.

It is intriguing that Ellison \etal\ (2006) found evidence of large dust depletion in intervening \MgII\ absorbers along GRB~060418. For one of the absorbers, at $z=1.11$, a 2200 \AA\ bump, typical in the Galactic extinction, was clearly detected for the first time at high redshift. This has yet to be found in QSO spectra.

Though most surveys suggest there is no selection bias associated with dust extinction in QSO spectra, this statement is valid for the total population of QSO absorbers, which are dominated by weak absorbers. And these do not affect the background light. Though the purpose of this survey is not to quantify the amount of dust extinction in QSO or GRB lines of sight, both the results obtained by P06 and us do not exclude these arguments. 

\subsection{Gravitational Lensing}

A similar selection bias can occur when the background source is magnified through gravitational lensing. The amount of magnification depends on the source beam size with respect to the Einstein radius of the lens, where the amplification is larger for a smaller source beam size (Chang \& Refsdal 1979). If it is indeed the case that GRB beams are smaller than QSO beams, gravitational magnification could be stronger for GRBs than for QSOs.

However, a larger amplification of GRBs is not enough (see for example Smette, Claeskens \&  Surdej 1997). The optical depth for lensing (macro or micro-lensing)  is maximal if the redshift of the lens is  $z_l\sim 0.7$ (it decreases to half the maximum  roughly  within $\Delta z =\pm 0.3$) for backgrounds sources (GRB or QSO) at $z_s \gg 1$. For background sources at $z_s \sim 1$ or less, the optical depth is maximal at $z_l \sim z_s/2$. Therefore, if lensing is really at play, then most \MgII\ absorbers would follow this. This is not the case of the \MgII\ absorbers. About 60\% (10 out of 16) of the GRBs studied by P06 and us are at $z_s>2$, and for these the foreground \MgII\ absorbers are mostly at $z_l>1$. For the remaining 40\% of the GRB sample, all but one have $z_s>1$ and $z_l>z_s/2$.

Porciani \etal\ (2007) tentatively found that afterglows with more than one \MgII\ absorber are on average 1.7 times brighter than their counterparts. This is based on the optical luminosity measured 12 hours after the burst (Nardini \etal\ 2006). We performed a similar check using the $B$-band absolute magnitudes calculated one day after the burst by Kann, Klose \& Zeh (2006) and Kann \& Klose (in preparation). This is available for 8 of the 14 GRB sight-lines in P06's \MgII\ sample. More appropriately, we considered the number density (number of \MgII\ absorbers per unit redshift) for each sight-line, instead of the total number of \MgII\ absorbers. We do not detect any correlation between the brightness of the GRB and the number density of \MgII\ absorbers. This test does not confirm the magnification hypothesis, though alone it cannot exclude it.

Porciani \etal\ (2007) suggest that up to 30\% of \emph{Swift} GRBs could be micro-lensed, making optical spectra easier to take. Micro-lensing could also give rise to time variability, following a scenario similar to the one considered by Lewis \& Ibata (2003), which would explain this effect found for \FeII\ and \MgII\ absorbers, by Hao \etal\ (2007).  Though a survey of \CIV\ would be less affected by macro-lensing due to foreground galaxies than a survey of \MgII , this is not the case for microlensing (microlenses are randomly distributed, whereas macrolensing becomes increasing significant with decreasing impact parameters). For this reason, it is unclear what significance these results have upon the hypothesis that gravitational lensing plays a role in the number density of \CIV .

\subsection{Absorbers associated with the GRB host environment}

It is generally accepted that metals that are ejected at semi-relativistic speeds from QSO or GRB sources would have distinctively broad profiles. These features can be found in QSOs up to $\Delta v \approx 50,000$ \kms\ away from the source, and are known as Broad Absorption lines  (BALs; Hamann et al.\ 1993). However, there has been no evidence that suggests that the same process occurs in GRBs. In fact, BALs are mostly detected in high-ionization absorption lines, which are naturally expected for gas clouds very close to a very luminous source. Therefore, if it is detected for \MgII\ absorbers, it should be even stronger for \CIV. Moreover, as relativistic speeds are expected, the absorptions features should be very large in the velocity space. This is clearly seen in BALs, but not at all present in normal intervening \MgII\ or \CIV\ absorbers.

The scenario proposed by Porciani \etal\ (2007), where a supernova remnant associated with the GRB and its environment, with multiple high velocity \MgII\ 'clouds', would give a \MgII\ number excess is not supported by our results. It is true that this effect might be less significant in \CIV\ absorbers, because the intrinsic number density of intervening systems is large and the contamination can only be small. Whereas,  there are generally few \MgII\ absorption systems per sight-line, this misidentification can significantly affect statistics. This scenario must only occur once or twice among the \MgII\ 14 GRB sample to significantly affect results. However, such high-velocity absorbers would require velocities of the gas of half the speed of light to be misidentified as an 'intervening absorber'. This is not typically observed in SN remnants. Intrinsic broadening of the absorbing features is also expected to be much larger than those observed.

 %%%%%%%%%%%%%%%%%%%%%%%%%%%%%%%%%%%

\section{Summary} %%SUMMARY

An excess of \MgII\ intervening absorbers along high-$z$ GRB sight-lines relative to similar absorbers found in high-$z$ QSO spectra was recently found. These \MgII\ absorbers trace galaxies intersecting bright background point sources. The total number of absorbers  in GRB sight-lines is nearly 4 times higher, in the interval $0.5<z<2$ (Prochter et al.\ 2006).

We have performed a similar study using an additional sample of \MgII\ absorbers, and confirm the previous results. We additionally investigate intervening \CIV\ absorbers and adopted a more detailed analysis. We have compared the number density and column density distribution in 3 GRB sight-lines at $1.6<z<3.1$ with the same quantities in QSO sight-lines in a similar redshift interval.  There is no detectable difference in the incidence of \CIV\ systems in the two samples, contrary to what found for \MgII.

Observationally, the main differences between the \CIV\ and \MgII\ absorbers are the larger impact parameter, the lesser dust extinction, the higher ionization level, the higher redshift, and the larger number density $dn/dz$ of the former with respect to the latter sample. We expect one or more of these factors to be directly or indirectly responsible for the discrepancy.

We exclude the possibilities that some of the intervening \MgII\ or \CIV\ absorbers are intrinsically associated with the GRBs, or that these absorbers have a partial covering factor in QSO spectra. We have shown that the scenario in which statistics being affected through the misidentification of intrinsic systems as intervening systems is unlikely, especially in a survey of \CIV\ absorbers.

Out of the possible effects, dust extinction could cause part of the difference in \MgII\ absorbers. This is due to the fact that QSO studies are generally performed on the bright population, whereas GRBs are primarily selected because of their gamma-ray signal. However, we have shown that dust gives a sizable effect only for the strong \MgII\ absorbers, those with $W_r>1.5$ \AA, which is half of the total sample used by P06. This is true for normal dust, however, with a dust-to-metals ratio similar to the Milky Way. If the dust-to-metals ratio is larger that what was previously thought for at least a fraction of high-$z$ galaxies, the effects of dust extinction would become more pronounced.

In the future, better statistics both for \CIV\ and \MgII\ absorbers in GRB spectra will help to solve the issue. \FeII\ absorbers could be used to probe the same galaxy regions as \MgII\ absorbers. The advantages of using \FeII\ are its larger redshift range (for a higher number statistics), and its less complicated access to column density. The column density is a more meaningful parameter than the equivalent width, used for \MgII. A further test is to study the mean Ly$\alpha$ absorption in the diffuse IGM, which is well studied from QSO spectra, and relatively easy to measure in GRB spectra, provided that statistical methods are used, such as the IGM optical method (Fan et al.\ 2006).

GRB afterglow studies have already demonstrated that there is a population of absorbers, identified as GRB-DLAs and intrinsic to the GRB host, that were very hard to detect before and that show on average very strong absorption features (Savaglio et al.\ 2003; Berger et al.\ 2006; Fynbo et al.\ 2006). It is not known whether the existence of this population is intrinsic to the GRB itself or common in the high-$z$ universe. The study of intervening absorbers presented in this work may show that GRBs are probing the high-$z$ universe better than QSO studies, regardless of the nature of the GRB itself. The results of our work are not conclusive, but they do imply that there may be fundamental assumptions about the structure of the universe and our methods to probe it that need to be reconsidered.

%%%%%%%%%%%%%%%%%%%%%%%%%%%%%%%%%%%
\acknowledgments %%ACKNOWLEDGMENTS

The authors thank Alec Boksenberg, Alex Kann, Cristiano Porciani and Tony Songaila for interesting comments. The authors especially thank Sara Ellison for sharing her insights concerning dust extinction and offering suggestions for follow-up work. VS acknowledges the generosity of the Winslow Womack grant, awarded by Guilford College for undergraduate research, and the kind hospitality of the Max-Planck Institute for Extraterrestrial Physics, during which a large part of this work was performed. 

%%%%%%%%%%%%%%%%%%%%%%%%%%%%%%%%%%%%

%%%%%%%%%%%%%%%%%%%%%%%%%%%%%%%%%%%%
%%FIGURES AND TABLES

%	\begin{figure}[htbp]%%%FIGURE: NUMBER DENSITY SMALL NCIV ONLY
%   	\begin{centering}
%    	\includegraphics[scale = .5]{NumberDensitySmallN.pdf}
%	\caption{\small The same quantities plotted as in Fig.~\ref{fig:NumberDensity}, but for the GRB and QSO \CIV\ subsamples with column density $N_{\rm C IV}\leq10^{13.8}$ cm$^{-2}$. In this case the redshift bin is $\Delta z=0.35$.}
%    	\label{fig:NumberDensitySmallN}
%	\end{centering}
%    	\end{figure}
%	
%	\begin{figure}[htbp]%%%FIGURE: NUMBER DENSITY LARGE NCIV ONLY
%   	\begin{centering}
%    	\includegraphics[scale = .5]{NumberDensityLargeN.pdf}
%	\caption{\small The same quantities plotted as in Fig.~\ref{fig:NumberDensity}, but for the GRB and QSO \CIV\ subsamples with column density $N_{\rm C IV}>10^{13.8}$ cm$^{-2}$. In this case the redshift bin is $\Delta z=0.35$.}
%    	\label{fig:NumberDensityLargeN}
%	\end{centering}
%    	\end{figure}
	
\begin{table} %%%TABLE: SAMPLE CHARACTERISTICS
\caption[t1]{Sample Characteristics}\label{table:samplecharacteristics}
\begin{center} 
\begin{tabular}{@{}lccccc@{}} 
\toprule
Source  & $z_{em}$ & $z_{min}-z_{max}$ & No. \CIV\ systems & FWHM 	      & S/N$^{\rm a}$\\ 
               &                    &                                    &					    & (km s$^{-1}$)  &	        \\
[3pt]\tableline\\
GRB 050820 	& 2.612	& 1.836-2.582	& 5	& 6.3		& 6 -- 18 --18 \\
GRB 050922C	& 2.198	& 1.511-2.168	& 7	& 6.8  	& 7-- 12 -- 12\\
GRB 060607	& 3.082	& 2.205-3.052	& 7	& 6.9 	& 58 -- n/a -- 61 \\
Q1626+6433	& 2.320     & 1.607-2.290	& 15	& 6.6 	& 88 -- 128 -- 137\\
Q1442+2931	& 2.661	& 1.875-2.631	& 14	& 6.6 	& 113 -- 107 -- 121\\
Q1107+4847	& 2.966	& 2.114-2.936	& 11	& 6.6 	&  90 -- 94 -- 81 \\
[3pt]\tableline
\end{tabular}
\end{center}
$^{\rm a}$ S/N over three pixels outside the Lyman forest sampled near the three wavelengths $\lambda \lambda \lambda$1270,1380,1500 in the rest frame of the GRB or QSO.
\end{table}

\begin{table} %%%TABLE: SYSTEM CHARACTERISTICS
\caption[t1]{\CIV\ Systems}\label{table:system_characteristics}
\footnotesize
\begin{center} 
\begin{tabular}{@{}ccccc@{}} 
\toprule
 &System & $z_{\rm abs}$ & $\log N_{CIV}$      & $b$ 		    \\ 
&	     &          &  $[cm^{-2}]$    &  (\kms) \\
\hline
GRB 050820 		& 				& 			&		&		\\
				& 1			 	& 2.0749 		& 13.64 	& 22.1	\\
				& 2 				& 2.1463		& 13.43	& 23.3	\\
				& 				&	  		& 13.33	& 6.1		\\
				& 3				& 2.3571		& 13.43	& 23.3	\\
				&				&			& 12.83	& 6.5		\\
				& 				&			& 13.21	& 5.9		\\
				& 4				& 2.3618		& 13.05	& 17.4	\\
				& 5				& 2.5706		& 13.12	& 14.5	\\
GRB 050922C		&				&			&		&		\\
				& 1				& 1.5670		& 13.78	& 14.7	\\
				& 2				& 1.5694		& 14.20	& 14.1	\\
				& 				& 			& 14.20	& 18.2	\\
				& 3				& 1.6911		& 14.44	& 25.3	\\
				& 				&			& 14.61	& 20.9	\\
				& 4				& 1.9891		& 14.26	& 13.1	\\
				& 				&			& 13.55	& 12.7	\\
				& 5				& 2.0087		& 13.60	& 15.7	\\
				& 				&			& 17.20$^{a}$	& 5.6		\\
				&				&			& 13.73	& 11.3	\\
				& 6 				& 2.0778		& 13.89 	& 17.9	\\
				& 7 				& 2.1421		& 13.47	& 5.1		\\
GRB 060607		&				&			&		&		\\
				& 1				& 2.2165		& 13.59	& 28.4	\\
				&				&			& 13.21	& 15.5	\\
				& 2				& 2.2180		& 13.55	& 6.6		\\
				& 3				& 2.2783		& 13.64	& 13.8	\\
				&				&			& 12.95	& 8.2		\\
				&				&			& 13.01	& 8.6		\\
				& 4				& 2.8903		& 14.12	& 5.6		\\
				&				&			& 14.27	& 11.7	\\
				&				&			& 13.75	& 7.8		\\
				&				&			& 13.95	& 8.0		\\
				&				&			& 13.42	& 9.8		\\
				& 5				& 2.9296		& 13.71	& 12.0	\\
				& 6				& 2.9352		& 14.14	& 27.4	\\
				&				&			& 14.54	& 28.8	\\
				&				&			& 14.68	& 35.7	\\
				& 7				& 3.0498		& 13.19	& 20.2	\\
				&				&			& 13.70	& 11.1	\\
\hline
$^{a}$ Heavily saturated. \\
\end{tabular}
\end{center}
\end{table}

\begin{table}[t] %%%TABLE: MgII SYSTEMS IN UVES SPECTRA
  \centering
  \caption[]{Identified \MgII\ systems in the UVES sample}
  \label{table:MgII_survey}
  \null\vspace{-0.5cm}
  $$
  \begin{array}{lllllc}
    \hline
    \hline
    \noalign{\smallskip}
    \rm GRB &
    z_{\rm GRB} &
    z_{\rm start} &
    z_{\rm end} &
    z_{\rm abs } &
    W_{r}(2796~\AA )^{\rm a} \\
    \hline
    050730	& 3.9686	& 1.160	& 2.0	& 1.7734 & 0.913 \pm 0.008 \\
    050820	& 2.6145	& 0.572	& 2.0	& 0.6915 & 2.896 \pm 0.017 \\
    		& 		& 	& 	& 1.4288 & 1.329 \pm 0.022 \\
    		& 		& 	& 	& 1.6204 & 0.238 \pm 0.009 \\
    050922C	& 2.1991	& 0.391	& 2.0	& 0.6369 & 0.167 \pm 0.009 \\
    		& 		& 	& 	& 1.1068 & 0.589 \pm 0.024 \\
    060418	& 1.4900	& 0.359	& 2.0	& 0.6026 & 1.257 \pm 0.008 \\
    		& 		& 	& 	& 0.6558 & 1.071 \pm 0.008 \\
    		& 		& 	& 	& 1.1069 & 1.836 \pm 0.006 \\
    060607	& 3.0748	& 0.772	& 2.0	& 1.5103 & 0.199 \pm 0.006 \\
     		& 		& 	& 	& 1.8033 & 1.906 \pm 0.008 \\
    \hline
  \end{array}
   $$\flushleft
  $^{\rm a}$ The error on the equivalent width is calculated from the noise spectrum; an
  error on the continuum normalization has not been included.\\
\end{table}

\end{document}